\documentclass[prb,twocolumn,preprintnumbers,amsmath,amssymb]{revtex4}
\usepackage{graphicx}
\usepackage{amssymb}
\usepackage[latin1]{inputenc}
\usepackage{dcolumn}% Align table columns on decimal point
\usepackage{bm}% bold math
\usepackage{amsmath}
\usepackage{mathtools}
\usepackage{textcomp}
\usepackage{amsbsy}
\usepackage{float}
\usepackage{times}
\DeclareGraphicsExtensions{.png,.eps}
\usepackage{xcolor}

\begin{document}

%\begin{CJK*}{GB}{gbsn}

%\draft

%\draft

\title{Surface States and Arcless Angles in Twisted Weyl Semimetals}

\author{Ganpathy Murthy$^{1}$, H.A. Fertig$^{2}$, and Efrat
  Shimshoni$^{3}$} \affiliation{$^{1}$ Department of Physics and
  Astronomy, University of Kentucky, Lexington, KY 40506-0055}
\affiliation{$^{2}$ Department of Physics, Indiana University,
  Bloomington, IN 47405} \affiliation{$^{3}$ Department of Physics,
  Bar-Ilan University, Ramat-Gan 52900, Israel}

\date{\today}

% buscar PACS
%\keywords{Topological insulators \sep Electronic properties \sep
%Transport properties}
%\pacs{73.20.At,75.70.Rf,75.30.Gw}

\begin{abstract}
Fermi arc states are features of Weyl semimetal (WSM) surfaces which
are robust due to the topological character of the bulk band
structure.  We demonstrate that Fermi arcs may undergo profound
restructurings when surfaces of different systems with a well-defined
twist angle are tunnel-coupled.  The twisted WSM interface supports a
moir\'e pattern which may be approximated as a periodic system with
large real-space unit cell.  States bound to the interface emerge,
with interesting consequences for the magneto-oscillations expected
when a magnetic field is applied perpendicular to the system surfaces.
As the twist angle passes through special ``arcless angles'', for
which open Fermi arc states are absent at the interface, Fermi loops
of states confined to the interface may break off, without connecting
to bulk states of the WSM.  We argue that such states have interesting
resonance signatures in the optical conductivity of the system in a
magnetic field perpendicular to the interface.
\end{abstract}
\maketitle
%\end{CJK*}

%\widetext
{\it Introduction} -- Weyl semimetals (WSM's) are three-dimensional
materials which host electronic structures with unusual, robust
topological properties.  Their bulk band structures contain an even
number of ``Weyl points,'' locations where the constant energy
surfaces shrink to a point, which act as sources or sinks of Berry's
flux through two-dimensional surfaces surrounding them
\cite{Jia_2016,Armitage_2018}.  Their presence in such materials leads
to a number of remarkable phenomena, including non-conservation of
currents associated with individual Weyl nodes in co-linear electric
and magnetic fields, and an associated negative longitudinal
magnetoresistance \cite{Zhang_2016,Jia_2016}.  WSM's can arise in
spin-orbit coupled systems with either broken inversion symmetry
\cite{Murakami_2007} or time-reversal symmetry
\cite{Wan_2011,Yang_2011,Xu_2011,Burkov_2011}. For the latter
anomalous Hall effects \cite{Jungwirth_2002,Fang_2003} are expected,
as well as non-local transport properties
\cite{Parameswaran_2014,Stern_2015,Zhang_2017}.

\begin{figure}[t]
%\centering
\includegraphics[width=4.3cm]{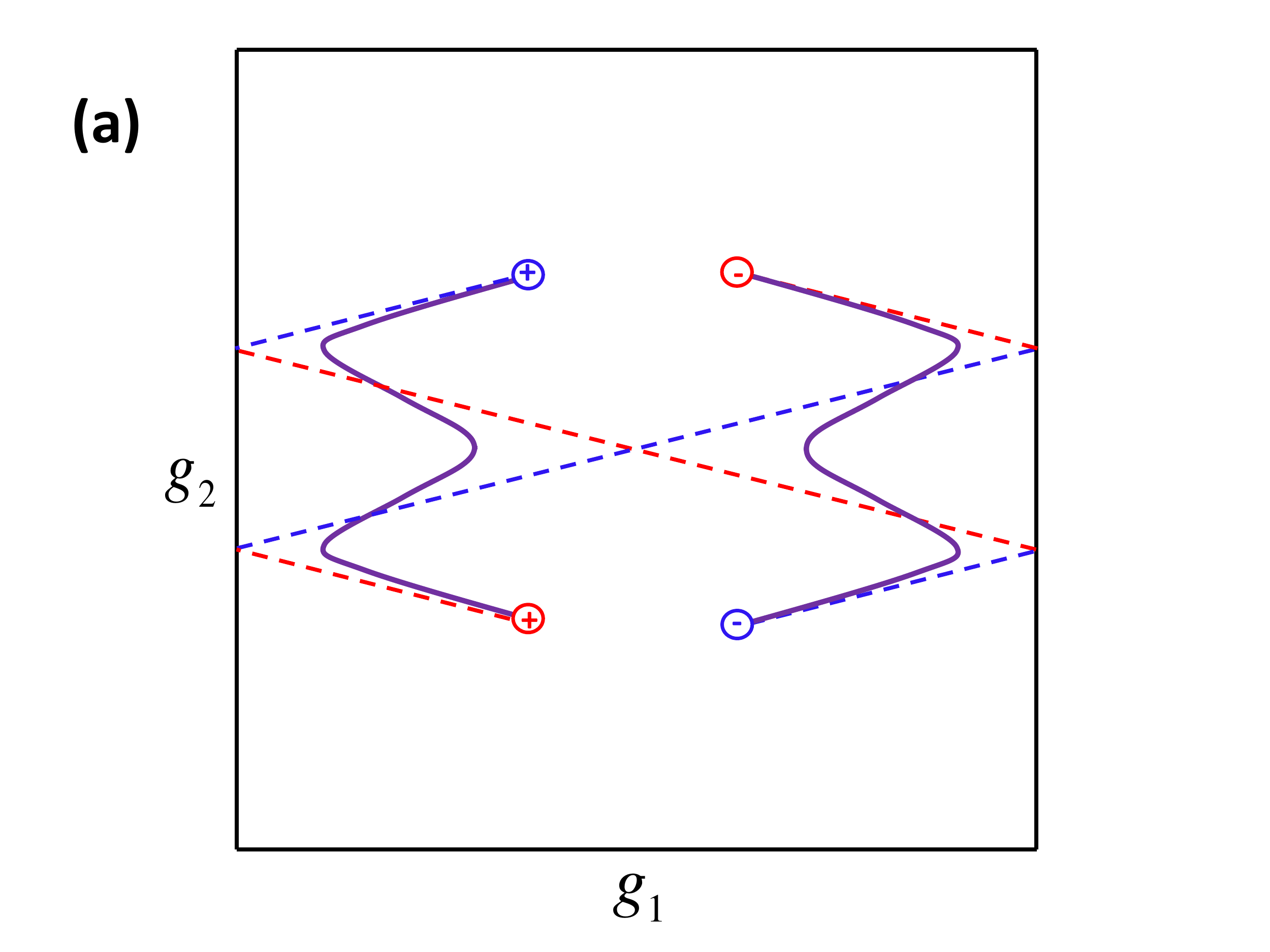}\!\!
\includegraphics[width=4.3cm]{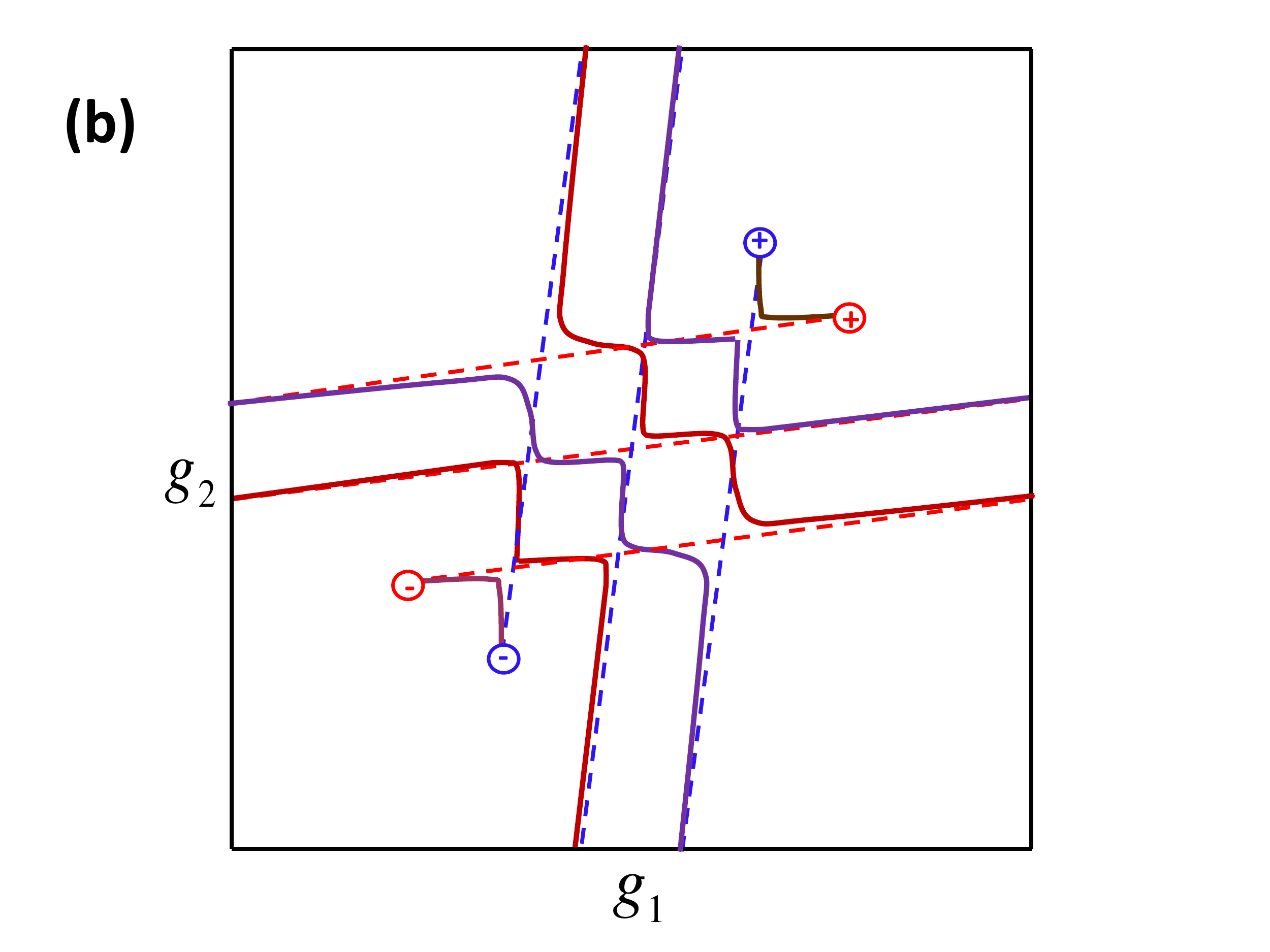}

\vspace{0.1cm}
\includegraphics[width=4.3cm]{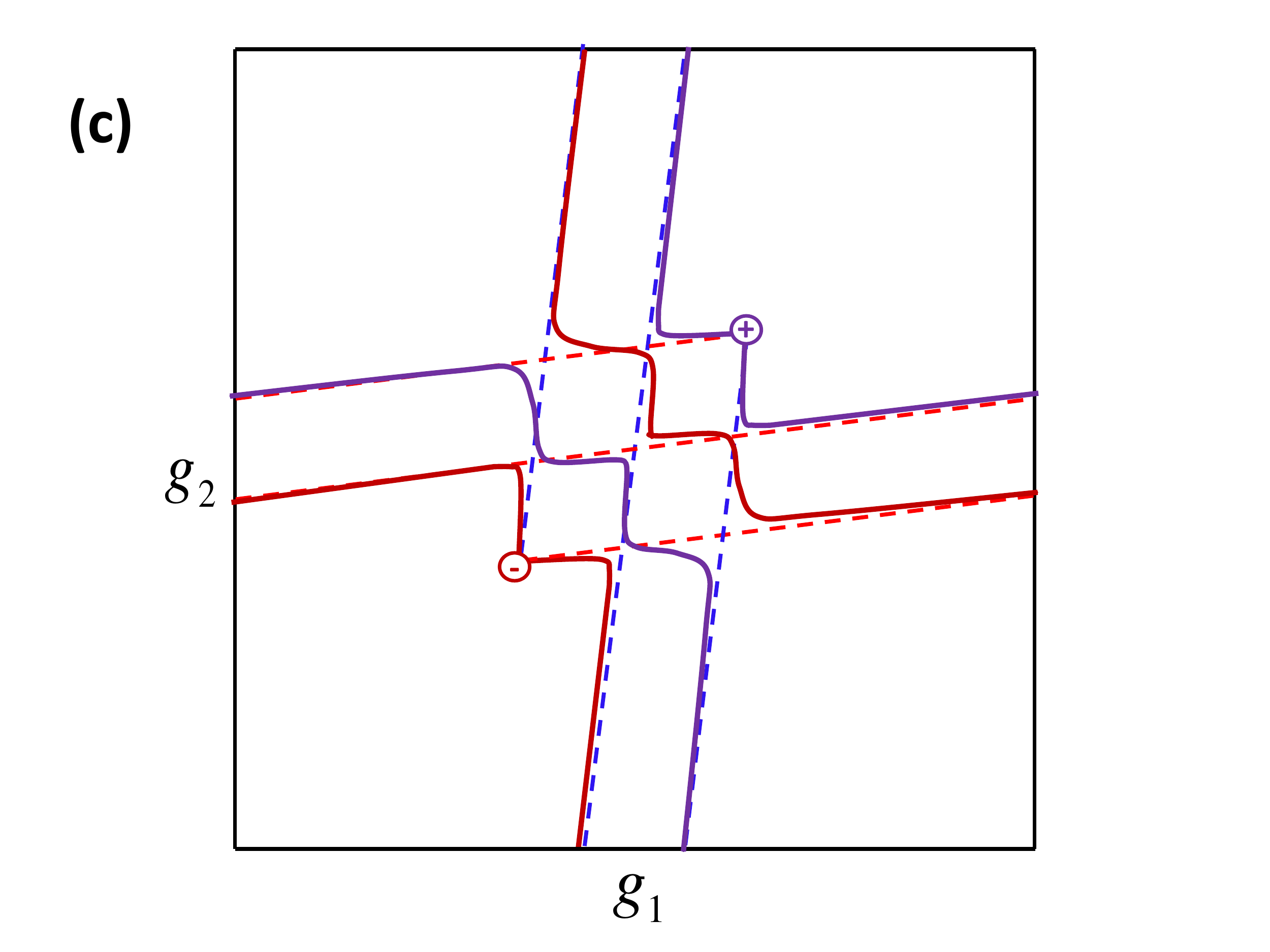}\!\!
\includegraphics[width=4.3cm]{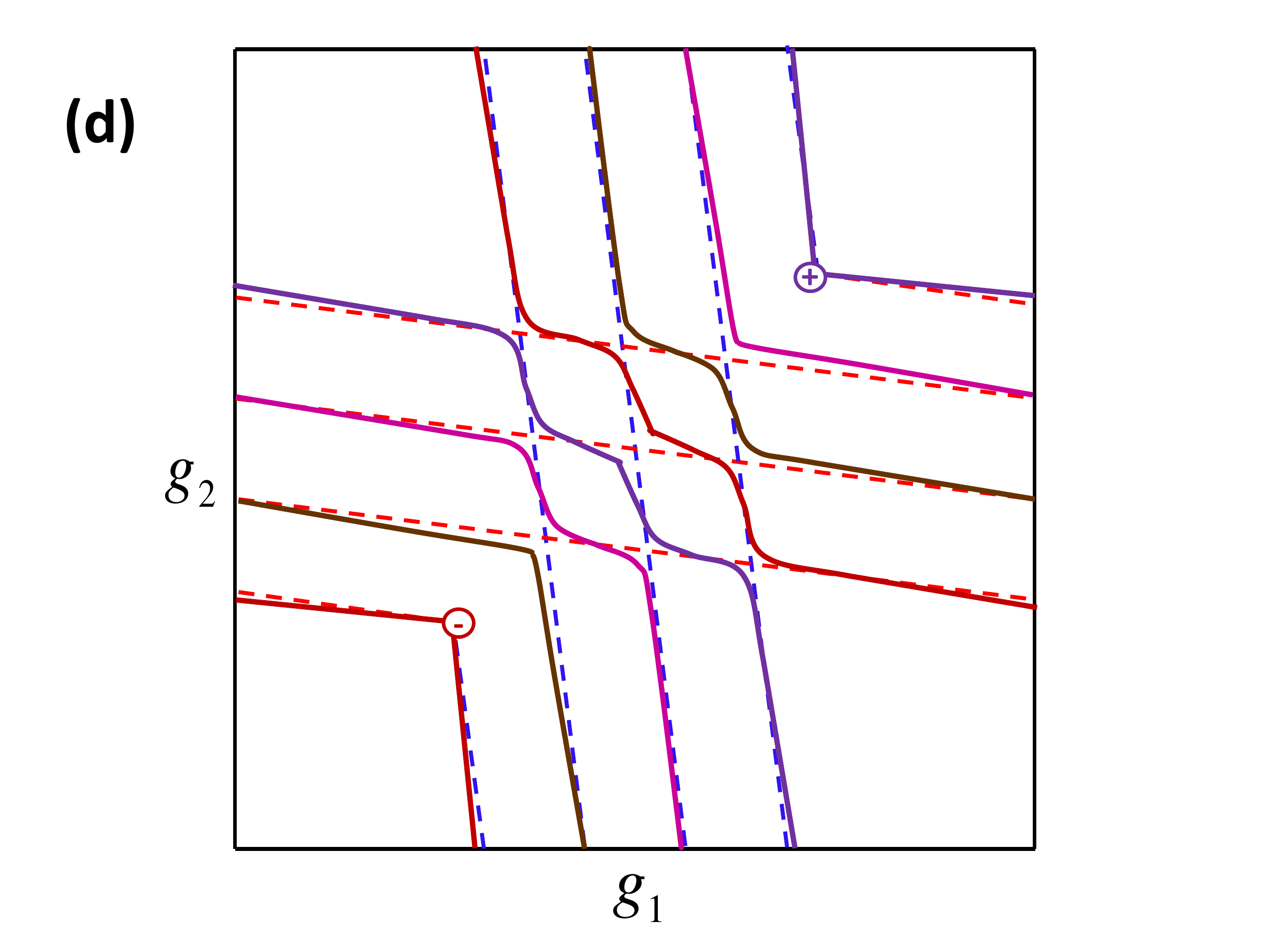}
\caption{Reconstructed Fermi arcs patterns in the moir\'e Brillouin
zone for various values of the twist angle $\theta$: (a) $\theta<\pi/4$; (b) an arbitrary $\pi/4<\theta<\pi/2$; (c) the arcless angle $\theta_{1}$, where two closed loops are generated; (d) the arcless angle $\theta_{-2}$, yielding two distinct pairs of closed loops (of which only one pair connects to the Weyl nodes). Dashed blue (red) lines represent the original arcs on the top (bottom) WSM slab. }
\label{fig:arc_reconstruct}
\end{figure}

Intimately connected with many of these phenomena is the presence of
``Fermi arc'' states \cite{Wan_2011,Ojanen_2013} in finite size WSM's.  For a
semi-infinite WSM, there is a two-dimensional (surface) Brillouin zone
(BZ) which generically hosts bound surface states.  At zero energy
(defined as the energy of the Weyl points), Fermi arcs are one or more
curves in the surface $\mathbf{k}$-space that connect the projections
of the Weyl points onto the surface BZ.  For any $\mathbf{k}$ that a
Fermi arc passes through, there is a state carrying surface current
perpendicular to the arc itself.  At the arc endpoints, the
penetration depth of the surface state diverges, allowing for exchange
between surface and bulk currents.  Indeed, in a slab placed in a
magnetic field perpendicular to the surfaces, current may flow in
opposite directions along Fermi arcs on each surface, connected by
chiral states through the bulk, to form closed loops
\cite{Potter_2014}.  These periodic orbits are expected to generate
magneto-oscillations in measurable properties of a
slab of WSM\cite{Moll_2016,Zhang_2019}.  Notably, because the orbits
involve a bulk component, the characteristic frequencies depend on the
slab thickness.

Given the remarkable nature of surface states in WSM's, it is
interesting to explore how their properties might be
modified and/or controlled.  This is the subject of our study.
Specifically, we consider the fate of Fermi arc states when surfaces
from two separate WSM's are twisted with respect to one
another, and then tunnel-coupled.  Tunnel-coupling of distinct Weyl
fermion systems aligned with one another along high-symmetry
directions are known to admit Fermi arc states which can connect Weyl
nodes of different subsystems \cite{Diwedi_2018,Ishida_2018}.  In this
work we consider general rotation angles, and find that the
possibilities for Fermi arc reconstruction are considerably richer.
This physics emerges from the lattice misalignment at the interface,
resulting in a moir\'e pattern that can be approximated by a
two-dimensional lattice with a large real-space unit cell.  Analogous
physics is known to be important in twisted graphene bilayers and
other two-dimensional van der Waals bonded systems
\cite{Novoselov_2016,Duong_2017}.  In the graphene case, the
prediction of extremely flat bands in the effective moir\'e Brillouin
zone \cite{Bistritzer_2011,SanJose_2012} has been indirectly verified
by the demonstration of interaction-induced correlated states
\cite{Cao_2018a,Cao_2018b} that one might expect of such systems.

Tunneling between states on the two WSM surfaces can occur directly
between states of the same crystal momentum (``direct tunneling''), or
between states whose wavevectors differ by linear combinations of the
reciprocal lattice vectors in each layer (``umklapp tunneling'').
When the layers are separated by a distance greater than the lattice
constant of the underlying crystals, the umklapp processes are
dominated by the shortest scattering wavevectors; retaining just the
two smallest of these (along with the direct tunneling term) results
in an effective surface superlattice defined by the two scattering
wavevectors, along with a corresponding moir\'e Brillouin zone.  We
find that, generically, the coupled Fermi arcs within a moir\'e
Brillouin zone will oscillate in direction (as shown in Fig. 1a), and typically connect Weyl
nodes of the two different slabs \cite{Diwedi_2018,Ishida_2018}.  For
some ranges of angles, one finds that Fermi arcs reconstruct to form
open arcs connecting the Weyl nodes, coexisting with arcs that close
upon themselves.  For special ``arcless angles'', the interface hosts
{\it only} closed Fermi loops.  For relatively weak coupling, some of
these loops include points at the surface projections of the Weyl
points, while for stronger coupling the Weyl points may couple
directly to one another, so that all the Fermi loops at the interface
are completely disconnected from the Weyl points.  A summary of these
different possibilities is illustrated in Fig. 1.

The existence of closed Fermi loops at the arcless angles is
particularly interesting.  For sufficiently clean
systems this means the interface hosts two types of conducting
channels.  Type (i) channels go through the Weyl point projections,
and have significant overlap with the bulk at the Weyl points. Type
(ii) channels never intersect the Weyl point projections, and thus
do not leak into the bulk of either system. In the presence of a
perpendicular magnetic field, currents
associated with type (ii) channels
should flow
along the arc.  Although they present open orbits in real space
-- and are thus not expected to lead to oscillations
in the density of states --
they involve periodic motion due to the
meandering of the orbits through the moir\'e Brillouin zone, with
periods inversely proportional to the magnetic field.  This suggests the
interface will present resonances in its optical conductivity
$\sigma(\omega)$ that reflect the orbit periods, in analogy with
the behavior of ``warped'' open orbits at the Fermi surface of
quasi-one-dimensional materials \cite{Ardavan_1998}.
Type (i) orbits similarly should present oscillations in $\sigma(\omega)$;
however, because they are {\it closed} in real space they are additionally expected to result
in density of states magneto-oscillations which
can be detected in thermodynamic quantities at zero frequency \cite{Potter_2014}.
In the
perturbative regime of tunnel-coupling, for type (i)
  orbits the resonance frequency in $\sigma(\omega)$ is expected to jump as the twist
  angle crosses an arcless angle. For type (ii) orbits, the frequency
  itself evolves continuously as a function of twist angle, but the
  amplitude of $\sigma(\omega)$ should jump as arcless angles
  are crossed.

We now turn to a more detailed discussion of our results, as well as the analysis that leads to them.

\textit{Model Hamiltonian and Surface States} -- Our starting point is
a simple two-band model \cite{Yang_2011} (with broken time-reversal
symmetry) of a WSM hosting two Weyl nodes, supporting a single Fermi
arc on any surface for which the projections of the Weyl points do not
overlap in the surface Brillouin zone.  The underlying crystal is
cubic (with lattice constant as our length unit), and the
bulk Hamiltonian as a function of crystal momentum takes the form
%and taking a surface perpendicular to the $\hat{z}$ direction, the Hamiltonian is
%\begin{widetext}
%\begin{eqnarray}
%H_0&=&
%\sum_{{\bf k},n}\sum_{s,s'} \psi_{{\bf k},n,s}^{\dag} \left[2 f_z({\bf k})\sigma_z +
%2f_x({\bf k})\sigma_x\right]_{s,s'} \psi_{{\bf k},n,s'} \nonumber \\
%&+&
%\sum_{{\bf k},n}\sum_{s,s'}\left\{ \psi_{{\bf k},n,s}^\dag %\left[-t\sigma_x+it'\sigma_y\right]_{s,s'}\psi_{n+1,s'} +
%\psi_{n+1,s}^\dag \left[-t\sigma_x-it'\sigma_y\right]_{s,s'}\psi_{{\bf k},n,s'} \right\}.
%\label{H0}
%\end{eqnarray}
%\end{widetext}
%$n$ labels a layer stacked in the
%${\hat z}$ direction, $s$ is a spin index, $\sigma_{x,y,z}$ are Pauli matrices, and $t$ and $t'$ are hopping matrix elements.  The functions $f_{x,z}$ are given by $f_x=t \sin k_y$ and
%
\begin{equation}
H_0({\bf k},k_z) = 2\sum_{\mu=x,y,z} f_{\mu}\sigma_{\mu}.
\label{H0}
\end{equation}
In this expression, ${\bf k}$ is a two-dimensional momentum,
$\sigma_{\mu}$ are Pauli matrices, and $f_x = t(2+\cos k_0 -\cos k_x -
\cos k_y- \cos k_z) \equiv t(1-\cos k_z) + \tilde f_x$, $f_y=t\sin
k_y$, and $f_z = t' \sin k_z $.  The bulk Weyl points are $(\pm
k_0,0,0)$. To simplify our notation, we hereon set $t=1$.  The
existence and form of surface states may be found by going to the
continuum limit in one direction, which we take here to be $\hat{z}$.
Expanding $H_0({\bf k},k_z)$ to second order in $k_z$, we can look for
evanescent solutions by taking $k_z \rightarrow i\lambda$.  Within
this procedure \cite{Liu_2010,Silvestrov_2012,Brey_2014}, four values
of $\lambda$ satisfying $H_0({\bf
  k},i\lambda)\mathbf{\Phi}=E\mathbf{\Phi}$ may be found, two of which
produce wave functions that vanish as $z \rightarrow \infty$.
%Note
%that the boundary conditions for the continuum and lattice approaches
%are different, as explained in the Supplemental Material (SM).
One
must then choose the value of $E$ such that the eigenvector
$\mathbf{\Phi}$ is the same for both these values of $\lambda$, so
that a linear combination may be formed satisfying vanishing boundary
conditions at $z=0$. (Details are provided in the SM, Sec. I.)  Three
points resulting from this analysis are particularly relevant to what
follows: (i) For the model defined by Eq. \ref{H0}, the procedure
described above can only be carried through for $|k_x| < k_0$.  Thus
the surface states are present on an open arc in the surface momentum
space.  (ii) Eigenstates of $H_0$ of this form satisfy $E=f_y(k_y)
\approx k_y$, so that the states have a velocity along the $\hat{y}$
direction. (iii) The eigenvectors that result from the analysis are
spin-polarized along the $\sigma_y$ direction.  For this simple model
we can also obtain surface states directly within the tight-binding model
(SM, Section II), yielding results consistent with the continuum
approximation. The lattice analysis has strengths complementary to the
continuum approach, in particular allowing the possibility of making the
tunnel-coupling nonperturbative, which we explore in detail in the SM,
Sections III and IV.

\textit{Twisted WSM's and Tunnel Coupled Fermi Arcs} --
Consider a pair of semi-infinite WSM's, one exposing a top surface and the other a bottom surface, each with a Fermi arc oriented at angles $\pm \theta$ with respect to the $\hat{x}$-axis.
For our model, the Fermi arc state dispersions
near zero energy can be written in the form $E_{T,B}({\bf k}) \approx {\bf v}_{T,B} \cdot {\bf k}$,
with ${\bf v}_T \cdot {\bf v}_b = -v_0^2 \cos 2\theta$.  In a long-wavelength model in which the
tunneling amplitude is uniform across the surfaces, the two-dimensional crystal momentum will
be conserved. Projecting into the Fermi arc states of the two surfaces, the coupled Hamiltonian may be modeled as
\begin{equation}
H_c=
\left(
\begin{array}{c c}
{\bf v}_T \cdot {\bf k} & w_d \\
w_d & {\bf v}_B \cdot {\bf k}
\end{array}
\right),
\label{Hc}
\end{equation}
where we have assumed $w_d$ is real.  In principle, after projection onto the surface states $w_d$ is a
function of ${\bf k}$, and should vanish at the positions of the arc end points, for which one
of the inverse length scales $\lambda$ in the surface state construction vanishes, indicating the
Fermi arc state crosses over into a bulk state. However, $w$ is most important near $k=0$ where the Fermi arcs cross, and is non-vanishing at that location.
Diagonalizing $H_c$ yields energies
$E_{\pm}=-v_0 \sin\theta k_x \pm \left[v_0^2\cos^2\theta k_y^2+ w_d^2\right]^{1/2}$, whose zero energy
contours are illustrated in Fig. 2(a).  Note that these reconstructed Fermi arcs now
connect Weyl nodes on opposite sides of $z=0$.

In reality, the two surfaces each have a square lattice structure, and
this makes the physics notably richer.  Primitive lattice vectors for
the top (bottom) surface may be written as ${\bf G}_1^{(')}$ and ${\bf
  G}_2^{(')}$ as illustrated in Fig. 2(b) for a small twist angle.  A
simple model incorporating this \cite{Bistritzer_2011} assumes that
the tunneling amplitude between atoms in different systems depends
only on the total distance between them; summing over the
two-dimensional direct lattice vectors then leads to tunneling
processes in which the in-plane wavevector scatters by a difference in
reciprocal lattice vectors, ${\bf g}_{ij} \equiv n_i{\bf G}_i -
n_j'{\bf G}_j'$, with $i,j=1,2$ and $n_i,n_j'$ integers.  Provided the
states are further apart than the underlying lattice constants of the
bulk materials, tunneling amplitudes will drop rapidly with increasing
$|n_i|$, $|n_j'|$.  The largest amplitudes for such umklapp tunneling
processes occur when one of these integers vanishes and the other is
$\pm 1$.  These processes may be regarded as renormalizations of the
direct tunneling process.  Qualitatively new behavior emerges when
$|n_i|=|n_j'|=1$, {\it and} translation of the Fermi arcs by ${\bf
  g}_{ij}$ causes them to overlap at new locations in momentum.

\begin{figure}[t]
\centering
\includegraphics[width=5.3cm]{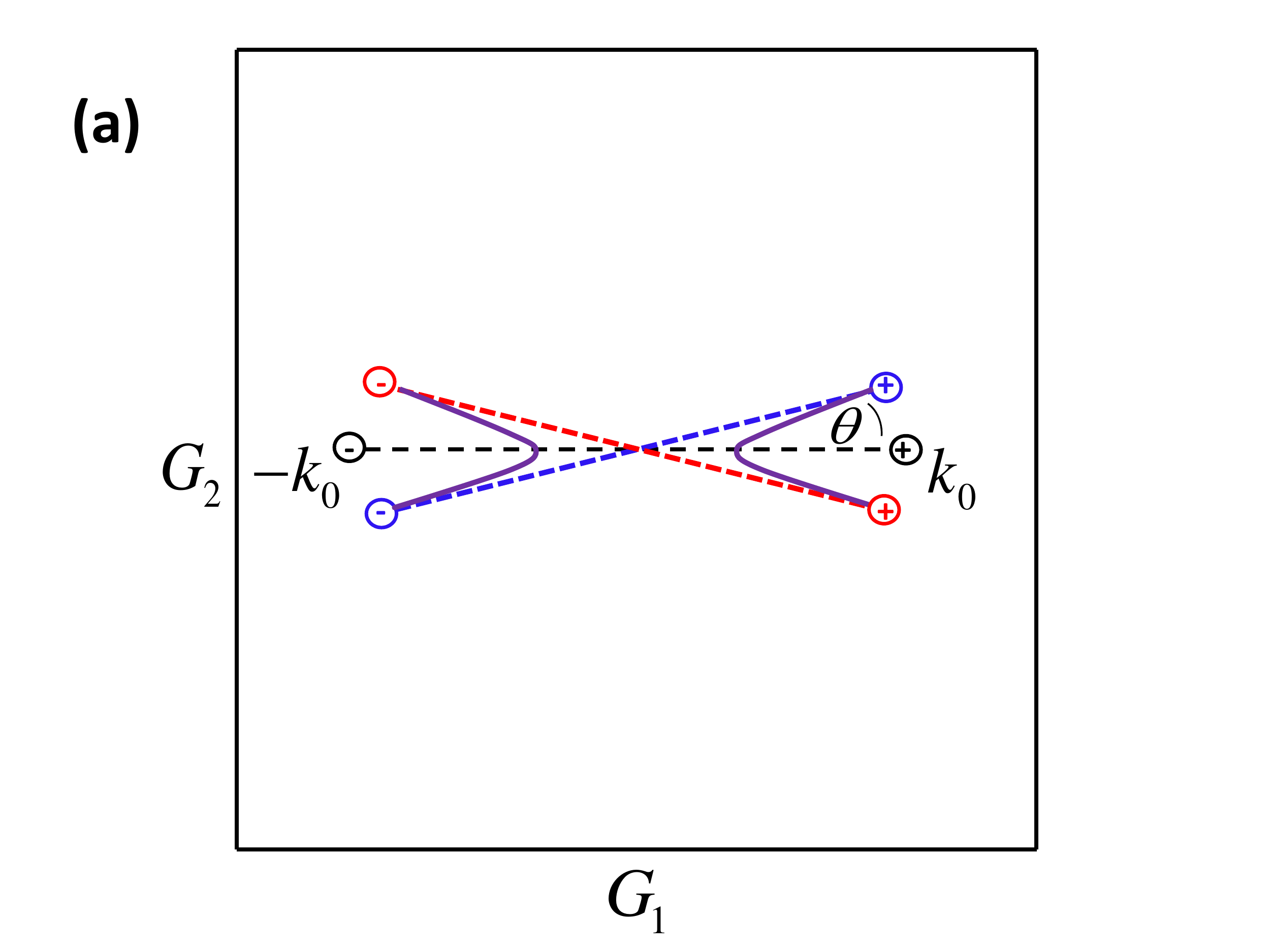}

\vspace{0.1cm}
\includegraphics[width=4.3cm]{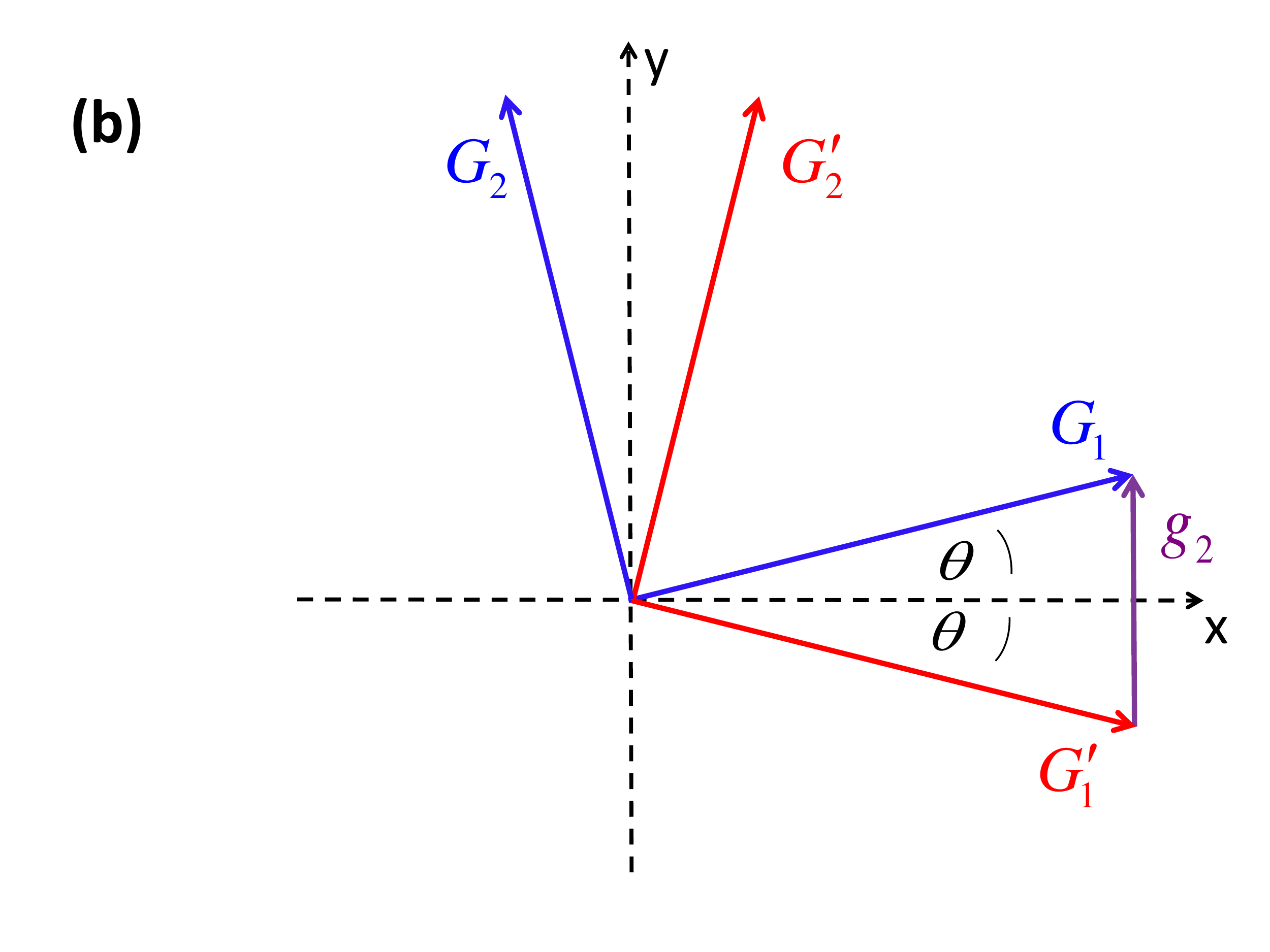}\!\!
\includegraphics[width=4.3cm]{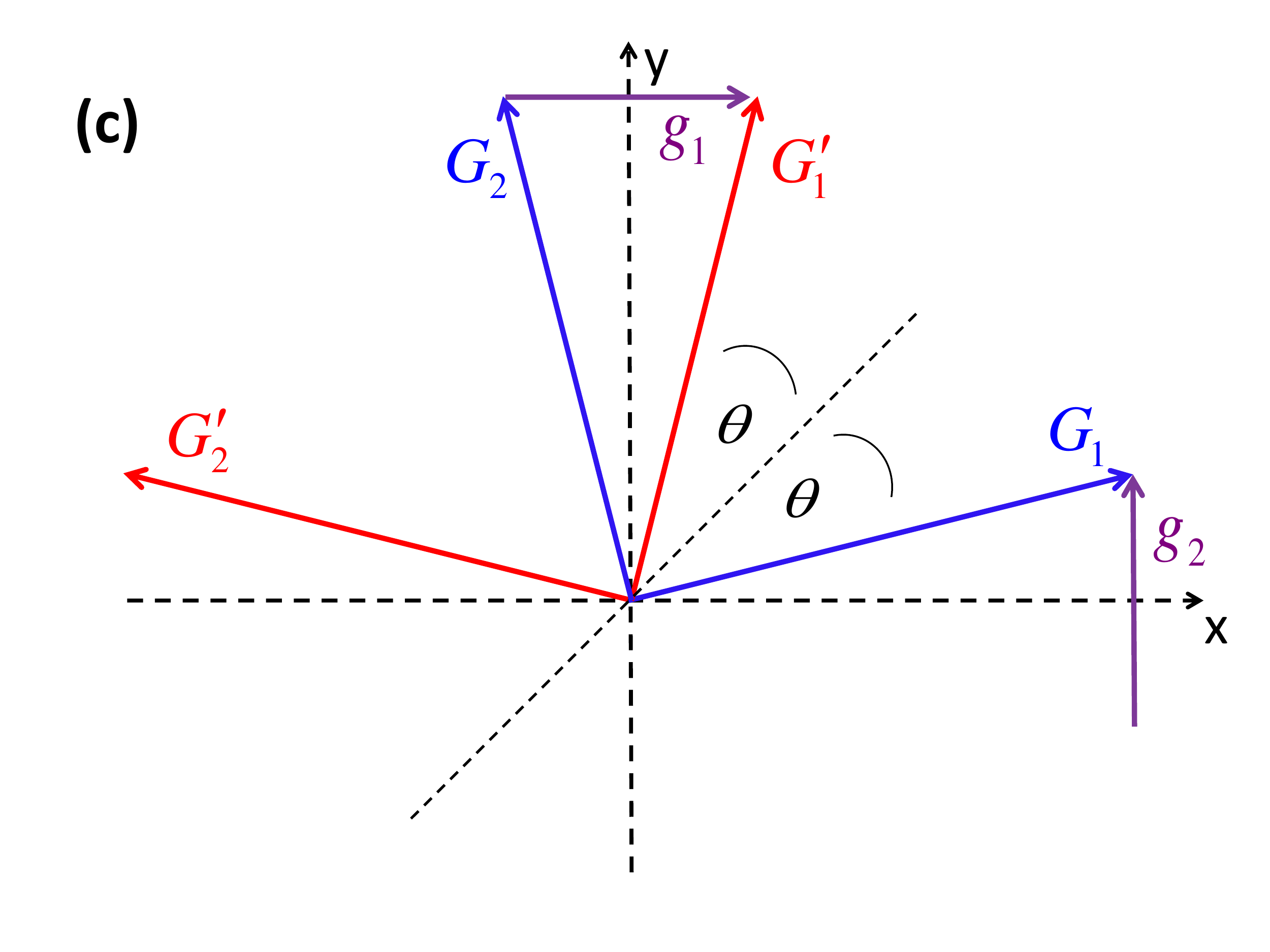}
\caption{(a) Zero energy contours (solid purple curves) resulting from the reconstruction of Fermi arcs (dashed lines) on the coupled surfaces of twisted WSM's. (b) Primitive lattice wavevectors of the top and bottom surfaces for a small twist angle ($\theta<\pi/4$). (c) Primitive lattice vectors of the top and bottom surfaces for a twist angle $\pi/4<\theta<\pi/2$; the wavevectors ${\bf g}_1$, ${\bf g}_2$ define the moir\'e Brillouin zone.}

\label{fig:MBZ}
\end{figure}

The Hamiltonian for states projected onto the Fermi arcs in this case takes the form
\begin{equation}
H_c=
\left(
\begin{array}{c c}
{\bf v}_T \cdot ({\bf k}+{\bf g}_{ij}) & w_u \\
w_u & {\bf v}_B \cdot {\bf k}
\end{array}
\right),
\label{Hc_um}
\end{equation}
where $w_u$ is a tunneling amplitude associated with the umklapp
process defined by ${\bf g}_{ij}$.  Within the assumptions discussed
above we expect $|w_u| \ll |w_d|$; however, for values of ${\bf k}$
such that ${\mathbf v}_T \cdot ({\mathbf k}+{\mathbf g}_{ij}) \approx {\mathbf v}_B
\cdot {\mathbf k}$ its effect cannot be neglected.  By retaining just the
vectors $\mathbf {g}_{ij}$ that generate such degeneracies, one can describe
the system in terms of an effective superlattice, which may be viewed
as an approximation to the moir\'e pattern \cite{Bistritzer_2011}
expected to appear at the interface between the two WSM's.  Using this
paradigm, we next turn to some of the interesting reconstructions of
the Fermi arcs that result from this physics.

\textit{Results} -- We begin by considering the situation for small
$\theta$.  In this case there is a single pair of collinear scattering
vectors, $\pm {\bf g}_1$ where $|{\bf g}_1|=2\sin\theta|{\bf G}_1|$
(see Fig. 2(b)), which are relevant to the Fermi arc restructuring.
In this case the system may be regarded as a one-dimensional
superlattice, and an example of the Fermi arc behavior is illustrated
in Fig. 1(a).  We assume the ``bare'' Fermi arcs to be
short compared to the surface Brillouin zone sizes of the individual
layers; nevertheless, for small enough $\theta$ the reconstructed
Fermi arcs oscillate back and forth repeatedly, ultimately connecting
a Weyl node of the lower system to one for the upper system.  As
$\theta$ increases the number of oscillations decreases, and
ultimately the reconstructed Fermi arcs will turn back once around
${\mathbf k}=0$ through a direct process to connect nodes in opposite
layers.

An interesting consequence of this behavior is its effect on
magneto-oscillations expected for WSM slabs \cite{Potter_2014}.  For
two rotated slabs tunnel-coupled in this way, in a magnetic field $B$
perpendicular to the interface, semiclassical orbits will pass through
the reconstructed arcs.  While the total arc length and the effective
Lorentz forces along most sections of the arcs are rather similar to
their values for uncoupled arcs, at the crossing points the Lorentz
force becomes $\sim B v_0 \sin \theta$, leading to a lengthening of
the total orbit period at small $\theta$, both because of the slow
velocity at an orbital turning point, and because of the proliferation
of the number of such turns as $\theta \rightarrow 0$.  This should
lead to an anomalous increase of the magneto-oscillation period at
small twist angles, which is eliminated as the twist angle increases,
both continuously as the velocity around the turning points increases,
and in steps as the ``switchbacks'' in the interface orbits are
eliminated.

Fig. 1(b) illustrates the situation for a twist angle satisfying
$2\theta \sim \pi/2$.  The relevant scattering wavevectors in this
case are $\pm{\bf g}_{1}$ and $\pm{\bf g}_{2}$, where ${\bf
  g}_1\equiv{\bf G'}_{1}-{\bf G}_2$, ${\bf g}_2\equiv{\bf G'}_{2}+{\bf
  G}_1$ [see Fig. 2(c)]. These wavevectors form an effective small
Brillouin zone $[{\bf g}_1,{\bf g}_2]$ that defines the wavevector
space for the periodic approximation to the moir\'e pattern at the
twist interface.  Because of the two-dimensional character of this
Brillouin zone, the pattern of avoided crossings breaks up the Fermi
arcs into a pair of short Fermi arcs connecting Weyl points in the two
different systems, and one or more longer closed loops
wrapping around the  torus defined by the moir\'e Brillouin
zone.  For specific twist angles, the Weyl nodes overlap perfectly,
eliminating the Fermi arcs, but leave behind the closed orbits (Fig 1c and 1d).  These
``arcless angles'' satisfy the equation
\begin{equation}
\tan \left(\frac{\pi}{2} - 2\theta_n \right)=\frac{k_0}{k_0+4\pi n},
\label{arcless}
\end{equation}
with $n$ an integer.  The situation at such angles is unique, in that
the twisted WSM supports $2|n|$ closed loops of surface
states. In the perturbative regime, two of these closed loops go through the Weyl
  point projections (type (i) loops), while $2(|n|-1)$ of
  them are composed of purely interface states (type (ii) loops).  The
number of such closed loops changes with twist angle, becoming
arbitrarily large as $2\theta \rightarrow \pi/2$. Note that at the
arcless angles, even the two type (i)  Fermi loops at the interface that
nominally overlap with the Weyl point projections may actually detach
from them at strong enough tunnel-coupling, as we demonstrate in
Sec. IV of the SM for a specific geometry. In such cases, {\it all}
surface states form closed loops disconnected from the Weyl point
projections. For small inter-surface coupling, it seems likely that
the type (i)  Fermi loops will contain the Weyl point projections.

Finally we note that for relative twist angles $2\theta \sim \pi$ and $2\theta \sim 3\pi/2$,
one expects to encounter restructured arc states with forms qualitatively similar to those
for small twist angles and for $2\theta \sim \pi/2$, respectively.

\textit{Discussion} -- The closed Fermi loops that form at twist
interfaces for relative angle $2\theta \sim \pi/2$ of type (ii), and
of both types at the arcless angles, represent one of the more
surprising phenomena occurring in the twisted WSM system.  A direct
probe of electrons in these states involves their semiclassical
dynamics in a magnetic field $\mathbf{B}$ perpendicular to the
interface.  The relevant equations of motion are
\begin{eqnarray}
\frac{d{\bf k}}{dt} &=& -e{\bf E}-{e \over c} {\bf v}_\alpha   \times {\bf B}, \nonumber \\
{\bf v}_\alpha({\bf k}) &=& \mathbf{\nabla} \varepsilon_\alpha({\bf k}).
\label{motion}
\end{eqnarray}
In these equations, ${\bf k}$ is contained in the moir\'e Brillouin
zone, and $\alpha$ labels different disconnected parts of the
reconstructed Fermi arcs. \cite{foot1} The energy function
$\varepsilon_\alpha({\bf k})$ corresponding to the closed Fermi arc is
given approximately, over most of its length, by the Fermi arc
energies for the uncoupled systems, appropriately translated into the
moir\'e Brillouin zone.  For trajectories near zero
energy at ${\bf E}=0$, and for $0 \le \theta \le \pi/4$ one should set
$\varepsilon_\alpha = E_-$ (with $E_-$ as defined below
Eq. (\ref{Hc})) near degeneracy points .

First we focus on type (ii) loops. The orbits of
Eqs. (\ref{motion}) are closed orbits in the moir\'e Brillouin zone,
but open in real space.  Magneto-oscillations in the electronic
density of states are not expected; however, the periodicity of the
$k$ space orbit should generate excitations of finite frequency,
detectable by resonances in the optical conductivity $\sigma(\omega)$
\cite{Ardavan_1998}.

As $\theta$ (assumed near $\frac{\pi}{2}$) decreases the period of the
type (ii) closed orbits in $\mathbf{k}$ decreases, and the frequency of the
resonance in $\sigma(\omega)$ increases smoothly as a function of
$\theta$. However, every time an arcless angle is crossed, the number
of type (ii) closed loops changes by two, which should lead to a jump in the
resonance {\it amplitude} in $\sigma(\omega)$. (These statements are
strictly correct for weak tunnel-coupling, for which all closed loops
are expected to have the same period for a given ${\mathbf B}$.)
Interestingly, because the orbits generating these resonances
oscillate both within the plane and perpendicular to them, they should
be detectable via absorption of electromagnetic waves polarized either
parallel or perpendicular to the interface \cite{Lu_2014}.

For generic angles, the type (i) interface orbits
that connect to the bulk Weyl points should admit more general
magneto-oscillations \cite{Potter_2014}, with period dependent on
length of the Fermi arcs at the interface, as well as those at the
more remote surfaces, which will also produce
  resonances in $\sigma(\omega)$. As the Fermi arcs abruptly change
length when $\theta$ passes through an arcless angle, we expect jumps
in the resonance frequency in $\sigma(\omega)$ due to
  type (i) orbits.

Finally, we briefly comment on the situation for WSM's where the Weyl
nodes appear due to broken inversion symmetry, rather than broken
time-reversal symmetry. A model bulk Hamiltonian of such a system, with four
bands, \cite{Vazifeh_2013} has the form \cite{Arijit_2019}
\begin{equation}\label{eq:H1}
H(k)=\lambda \sum_{\mu =x,y,z} \sigma_\mu \sin{k_\mu} + \tau_y \sigma_y M_k,
\end{equation}
where $M_k=m+2-\cos{k_x}-\cos{k_z} $, $ \tau_\mu $ are Pauli matrices
acting in an orbital space and $\sigma_\mu$ acts in a spin space.  For
this model we do not find situations in which closed Fermi loops form
at the interface which are disconnected from the Weyl nodes.
Nevertheless, three-dimensional closed orbits should still form in a
magnetic field, which pass through the interface states.  At small
twist angles some of these loops will acquire an anomalously long
period in the same way as was found above in the system with broken
time-reversal symmetry.  We again expect this to lead to
magneto-oscillations of suppressed frequency.

In real materials, surfaces can have significantly larger numbers of
Fermi arcs, which may connect amongst one another along arcs of
non-vanishing curvature.  At small twist angles we expect such systems
will also host low frequency magneto-oscillations.  It is interesting
to speculate that some such systems might also host closed Fermi loops
as found in the two-band model, as well the elimination of open Fermi
arcs connecting some of the bulk Weyl nodes.  We leave these questions
for future research.

\textit{Acknowledgements --} This work was supported by the US-Israel
Binational Science Foundation (Grant No. 2016130: GM, HAF, ES; Grant No. 2018726: HAF, ES), by the
NSF (Grant Nos. DMR-1506263, DMR-1506460, DMR-1914451), and by
the Israel Science Foundation (ISF) Grant No.~231/14 (ES).  The authors acknowledge the hospitality and
support of the Aspen Center for Physics (Grant No. PHY-1607611), where
part of this work was done. GM is grateful to the Gordon and Betty
Moore Foundation for sabbatical support at MIT, and the Lady Davis
Foundation for sabbatical support at the Technion, while these ideas
were being formed.

\newpage

\begin{widetext}

\section{Supplemental Material: Surface States and Arcless Angles in Twisted Weyl Semimetals}

Our Supplemental Material is organized as follows.
We start our discussion with the continuum model in Section
\ref{continuum}, which has the advantage of being easy to generalize
to the case of tunnel-coupled surfaces of Weyl semi-metals twisted
with respect to each other at arbitrary angle. The disadvantage is that one can only
incorporate tunnel coupling between the surfaces perturbatively.
In Section \ref{lattice-oneslab} we then develop a lattice formulation for the case of a single
semi-infinite slab.
We proceed further in
Section \ref{lattice-twoslabs} by considering the case of two slabs of
identical Weyl semi-metal twisted with respect to each other by
$\frac{\pi}{2}$ and tunnel-coupled to one another. The advantage of the
lattice formulation is that for twist angles that are multiples of $\pi/2$, it allows
a treatment that is nonperturbative in the tunnel-coupling.
Finally,
in Section \ref{curved-detach} we modify the Hamiltonian of
Eq. (\ref{eq:exact_H}) to obtain curved Fermi arcs, and show
by a lattice calculation that when the tunnel-coupling is sufficiently
strong, the Fermi arcs can detach from the projections of the Weyl
points, and form closed loops in the surface BZ.

\subsection{Fermi arc states for a single semi-infinite slab: Continuum approach}
\label{continuum}

We start with our tight-binding Hamiltonian of Eq. (1) in the main text, with $t$ set equal to unity from the beginning (and $t'>0$):
\begin{equation}
  H=2\sum\limits_{\mu} f_{\mu} \sigma_{\mu},
  \label{eq:exact_H}
\end{equation}
where $  f_x=2+\cos(k_0)-\cos(k_x)-\cos(k_y)-\cos(k_z)\equiv 1-\cos(k_z)+\tilde{f}_x$, $f_y=\sin(k_y)$, and $f_z=t'\sin(k_z)$.  (A special case of this model, $t'=1$, was considered in Ref. \onlinecite{Diwedi_2018}.)
We then employ a small $k_z$ expansion, yielding a continuum approximation in the $z$ direction (where the coordinate $z$ is measured in units of the lattice constant $a=1$). The functions $f_x$, $f_z$ now become
\begin{eqnarray}
  f_x&=&\frac{1}{2}k_z^2+\tilde{f}_x\; ,\quad \tilde{f}_x = 1+\cos(k_0)-\cos(k_x)-\cos(k_y),\nonumber\\
  f_z&=&t'k_z .
  \label{eq:def_f}
\end{eqnarray}
In a real-space representation of the $z$ direction, the Hamiltonian density acquires the form
\begin{equation}
  H(k_x,k_y,z)=-2it'\sigma_z\frac{\partial}{\partial z}+\left(2\tilde{f}_x-\frac{\partial^2}{\partial z^2}\right)\sigma_x+2f_y\sigma_y\; .
  \label{eq:H_z}
\end{equation}
Considering a semi-infinite slab where $z\geq 0$, we then search for eigenvectors $\mathbf{\Phi}(z)$ of $H$ (with eigenvalues $E(k_x,k_y)$) which obey vanishing boundary conditions on the surface $z=0$ and decay in the bulk $z\rightarrow\infty$.

From Eq. (\ref{eq:H_z}) it is apparent that $\mathbf{\Phi}(z)$ are linear combinations of eigenvectors of the form
\begin{equation}
\mathbf{\Phi}_\lambda(z)=\mathbf{\Phi_0}e^{-\lambda z}\; ,
\label{eq:Phi_lambda}
\end{equation}
where $\Re\{\lambda\}>0$ and $\mathbf{\Phi_0}$ is a ($z$-independent) two-component spinor. Substituting $\mathbf{\Phi}_\lambda$ in $H\mathbf{\Phi}_\lambda=E\mathbf{\Phi}_\lambda$, we obtain a quadratic equation for $\lambda$ as a function of $E,k_x,k_y$ with the solutions
\begin{equation}
  \lambda^2_\pm=2\left[\tilde{f}_x+(t')^2\right]\pm \sqrt{4\left[\tilde{f}_x+(t')^2\right]^2+E^2-4\left[\tilde{f}_x^2+f_y^2\right]}\; .
  \label{eq:lambda_vs_E}
\end{equation}
The corresponding eigenvectors are of the form Eq. (\ref{eq:Phi_lambda}) with
\begin{equation}
  \mathbf{\Phi_0}^{(\pm)}\equiv \left(\begin{array}{c}
    u_{_\pm}\\
    v_{_\pm}
  \end{array}\right)=\left(\begin{array}{c}
    2\tilde{f}_x-2if_y-\lambda^2_\pm\\
    E-2it'\lambda_\pm
  \end{array}\right)\; .
\label{eq:eigenvectors}
\end{equation}
The resulting eigenstates $\mathbf{\Phi}(z)=a_{_+}\mathbf{\Phi}_{\lambda_{+}}(z)+a_{_-}\mathbf{\Phi}_{\lambda_{-}}(z)$ obey the boundary condition $\mathbf{\Phi}(0)=0$ provided the coefficients $a_{_\pm}$ satisfy
\begin{equation}
  M\left(\begin{array}{c}
    a_{_+}\\
    a_{_-}
  \end{array}\right)=0\quad {\rm with} \quad M=\left(  \begin{array}{cc}
    2\tilde{f}_x-2if_y-\lambda^2_+ \;&\;2\tilde{f}_x-2if_y-\lambda^2_-\\
    E-2it'\lambda_+&E-2it'\lambda_-
  \end{array}\right)\; ,
\end{equation}
which has a solution provided $\det M=0$. Assuming $\lambda_+\neq\lambda_-$, this yields the equation
\begin{equation}
-E(\lambda_+ + \lambda_-)+4it'(\tilde{f}_x-if_y)+2it'\lambda_+\lambda_- \; .
\end{equation}
Assuming that $\lambda_+ , \lambda_-$ are real, they must obey simultaneously
\begin{equation}
-E(\lambda_+ + \lambda_-)+4t'f_y=0 \quad\quad {\rm and} \quad\quad 2\tilde{f}_x=-\lambda_+\lambda_- \; .
\label{eq:E2lambda}
\end{equation}

We next impose the requirement $\lambda_+ , \lambda_->0$, which
implies that the second conditions in Eq. (\ref{eq:E2lambda}) can be
satisfied only provided $\tilde{f}_x<0$. Additionally, for the special
case $E=0$ the first condition yields $f_y=0$. Recalling the
definition of $\tilde{f}_x$, $f_y$ [Eq. (\ref{eq:def_f})], we conclude
that a consistent solution for the surface states exists along the
segment $k_y=0$, $|k_x|<k_0$. Notably, this is precisely the Fermi arc
connecting the two Weyl nodes $(\pm k_0,0)$. For arbitrary $E$, we
combine Eqs. (\ref{eq:lambda_vs_E}) and (\ref{eq:E2lambda}) to get
\begin{eqnarray}
  E&=&2f_y\; ,\nonumber\\
  \lambda_\pm &=&t'\pm \sqrt{(t')^2+2\tilde{f}_x}\; .
  \label{eq:lambda_final}
\end{eqnarray}
Interestingly, when substituted in Eq. (\ref{eq:eigenvectors}) to obtain the eigenvectors, we find
\begin{equation}
\left(\begin{array}{c}
    u_{_+}\\
    v_{_+}
  \end{array}\right)=\left(\begin{array}{c}
    u_{_-}\\
    v_{_-}
  \end{array}\right)=\frac{1}{\sqrt{2}}\left(\begin{array}{c}
    1\\
    i
  \end{array}\right) \; ;
\end{equation}
i.e., the surface states are spin-polarized along the $\sigma_y$ direction, as stated in the main text.

\subsection{Fermi arc states for a single semi-infinite slab: Lattice calculation}
\label{lattice-oneslab}

In this section we consider the lattice Hamiltonian Eq. (\ref{eq:exact_H}), and
transform to real space in the $z$-direction, with the lattice
sites in this direction labeled by $n$.  The momenta $k_x,\ k_y$ will
be left as parameters. Defining the lattice Fourier transform by
\begin{eqnarray}
  c_n(k_x,k_y)=\frac{1}{\sqrt{N_z}}\sum\limits_{k_z} e^{ik_zna} c(k_x,k_y,k_z),\\
  c(k_x,k_y,k_z)=\frac{1}{\sqrt{N_z}}\sum\limits_{n} e^{-ik_zna} c_n(k_x,k_y),
\end{eqnarray}
where the lattice spacing is $a\equiv 1$ and we have suppressed the
spin index in the fermion destruction operators $c(k_x,k_y,k_z)$, $c_n(k_x,k_y)$. After a bit of
algebra we obtain
\begin{equation}
  H(k_x,k_y)=\sum\limits_{n} \bigg( -c^{\dagger}_{n+1} \big( \sigma_x-it'\sigma_z) c_{n}-c^{\dagger}_{n} \big( \sigma_x+it'\sigma_z) c_{n+1} +2c^{\dagger}_{n} \big((1+\tilde{f}_x)\sigma_x+f_y\sigma_y\big)c_{n}\bigg).
\label{real-spaceH}
\end{equation}
In the above we have suppressed the $k_x,\ k_y$
arguments in all fermion operators for notational simplicity. For
further analysis, it is convenient to rotate the $\sigma$-matrices by
$\frac{\pi}{2}$ around the $x$-axis, which has the virtue of making
the Hamiltonian purely real. In this basis, the Hamiltonian is
\begin{equation}
  H(k_x,k_y)=\sum\limits_{n} \bigg( -c^{\dagger}_{n+1} \big( \sigma_x+it'\sigma_y) c_{n}-c^{\dagger}_{n} \big( \sigma_x-it'\sigma_y) c_{n+1} +2c^{\dagger}_{n} \big((1+\tilde{f}_x)\sigma_x+f_y\sigma_z\big)c_{n}\bigg).
  \label{real-real-H}
\end{equation}
Since the Hamiltonian is real, the wavefunction can also be chosen
real. This will play an important role in the counting argument that
shows that the conditions to have a surface state can be met.

To consider a semi-infinite slab, we
simply restrict the sum over $n$ to either non-negative or
non-positive integers. For specificity, let us consider the bottom
surface of a semi-infinite slab, with $n=0,1,2,\dots$. We want
solutions that decay into the bulk, so we make the eigenstate ansatz
\begin{equation}
  |E,k_x,k_y\rangle=\sum\limits_{n=0}^{\infty}
  \sum\limits_{s=\uparrow,\downarrow}u^n\Phi_{0s}|n,s\rangle,
\label{decay-ansatz-b}
\end{equation}
where $|u|<1$ for the state to be normalizable. Demanding
$H|E,k_x,k_y\rangle=E|E,k_x,k_y\rangle$ leads to the matrix condition on $\Phi_{0s}$, for $n>0$,
\begin{equation}
  \bigg(u^2\big(\sigma_x-it'\sigma_y\big)+u\big(E-2(1+\tilde{f}_x)\sigma_x-2\sin(k_y)\sigma_z\big)+\big)\sigma_x+it'\sigma_y\big)\bigg)\Phi_0\equiv M\Phi_0=0.
\label{bulk-condition}
\end{equation}
In explicit terms, the matrix $M$ is
\begin{equation}
 M= \left(  \begin{array}{cc}
    u(E-2f_y)&1+u^2-2(1+\tilde{f}_x)+t'(1-u^2)\\
    1+u^2-2u(1+\tilde{f}_x)-t'(1-u^2)&u(E+2f_y)
  \end{array}\right).
\end{equation}
Since the matrix $M$ has a zero eigenvector, its determinant should
vanish. This gives a quartic equation for $u$, which, by changing to
the variable
\begin{equation}
  \xi=u+\frac{1}{u}
\end{equation}
can be recast as a quadratic equation,
\begin{equation}
  \xi^2(1-t'^2)-4\xi(1+\tilde{f}_x)-\big(E^2-4(1+\tilde{f}_x)^2-4f_y^2-4t'^2\big)=0.
  \label{xi-quadratic}
\end{equation}
Clearly, we can solve this quadratic and then the auxiliary quadratic
$u^2-u\xi+1=0$ to obtain four values of $u$. The structure of this
auxiliary quadratic shows that the values of $u$ occur in pairs whose
product is unity. Thus, either both roots are real, with one being
less than unity and the other greater, or they are unimodular and
complex conjugates of each other. In the latter case the wavefunctions
will not be normalizable, and thus there will be no surface modes.

It is interesting to consider the relation of our approach to that of Ref. \onlinecite{Diwedi_2018},
which is restricted to $t'=1$. Clearly, $t'=1$ is singular in the
sense that the quadratic equation Eq. (\ref{xi-quadratic}) becomes a
linear one. This means one of the roots has been pushed off to
infinity. Solving the auxiliary quadratic, we see that there are only
two physically acceptable roots for $u$ in this case, with only one of
them being less than unity in magnitude. Ref. \onlinecite{Diwedi_2018}
presents many elegant
results in this simplified model, but one might
worry that the results of the fine-tuned $t'=1$ model may not be
generic.  Our analysis demonstrates that in fact they are.

Going back to our problem, let us assume that we have obtained two
real values of $u$, say $u_1,u_2$ such that $|u_1|<1,\ |u_2|<1$. The
next step is to find the eigenvectors corresponding to these values of
$u_i$ by solving $M(u_i)\Phi_{0}^{i} =0$. The final step is to find a
linear combination of these solutions
\begin{equation}
  \Phi(n)=\alpha_1 u_1^n\Phi_0^1 +\alpha_2 u_2^n\Phi_0^2
\end{equation}
that satisfies the boundary condition on the surface, which can be
written as
\begin{equation}
  \Phi(n=-1)=0.
  \label{Phi-1eq0}
\end{equation}

Since the $u_i$ as well as the wavefunctions are real, so are
$\alpha_i$. The normalization is immaterial for satisfying
Eq. (\ref{Phi-1eq0}), which means we can set $\alpha_1=1$. Thus, for a
given $k_x,k_y$, we have one real free parameter $\alpha_2$, and
another real free parameter $E$. With these two parameters we can
indeed satisfy the two conditions represented by
Eq. (\ref{Phi-1eq0}). This means a solution exists as long as we can
find two real $u{(b)}_i$, with $|u^{(b)}_i|<1$. [Note that the superscript $(b)$
denotes the bottom surface.]

Given a $u$ which solves the characteristic equation, we can find the
eigenvector corresponding to it by solving
Eq. (\ref{bulk-condition}). For future reference, the un-normalized
spinor can be expressed in two equivalent ways,
\begin{eqnarray}
  \Phi_0(u,E,k_x,k_y)=&\left(\begin{array}{c}
    1\\
    -\frac{u(E-2f_y)}{u^2(1+t')-2u{F_x}+1-t'}
  \end{array}\right),  \nonumber \\{\rm or} \nonumber \\
  \Phi_0(u,E,k_x,k_y)=&\left(\begin{array}{c}
    -\frac{u(E+2f_y)}{u^2(1-t')-2u{F_x}+1+t'} \nonumber \\
    1
  \end{array}\right).
  \label{spinors}
\end{eqnarray}

The solutions turn out to be particularly simple if one assumes
$0<t'=\sin{\phi}<1$, and $k_0<\phi$.  For the bottom surface of a
semi-infinite slab, the energy is
\begin{equation}
  E^{(b)}(k_x,k_y)=2f_y=2\sin{k_y}.
  \label{Ebottom}
\end{equation}
(Note this same result was obtained previously \cite{Diwedi_2018} for the specific case $t'=1$.)  Defining
${F_x}=1+\tilde{f}_x=\cos{k_0}+\frac{1}{2}\big(\sin^2{\frac{k_x}{2}}+\sin^2{\frac{k_y}{2}}\big)$,
we find that the two $u_i$ we want are solutions to the vanishing of
the lower left entry of the matrix $M$.  Thus
\begin{equation}
  1+u^2-2{F_x}u-t'(1-u^2)=u^2(1+\sin{\phi})-2{F_x}u+(1-\sin{\phi})=0,
  \nonumber
\end{equation}
yielding the solutions
\begin{equation}
  u^{(b)}_{\pm}=\frac{{F_x}\pm\sqrt{{F_x}^2-\cos^2{\phi}}}{1+\sin{\phi}}.
  \label{eq:u_pm}
\end{equation}
The Fermi arcs persist within a region of the surface BZ where $  {F_x} <1$, or equivalently
\begin{equation}
\cos{k_x}+\cos{k_y}>1+\cos{k_0}
\label{FA-region}
\end{equation}
\begin{figure}
\centerline{\includegraphics[angle=0,width=0.4\columnwidth]{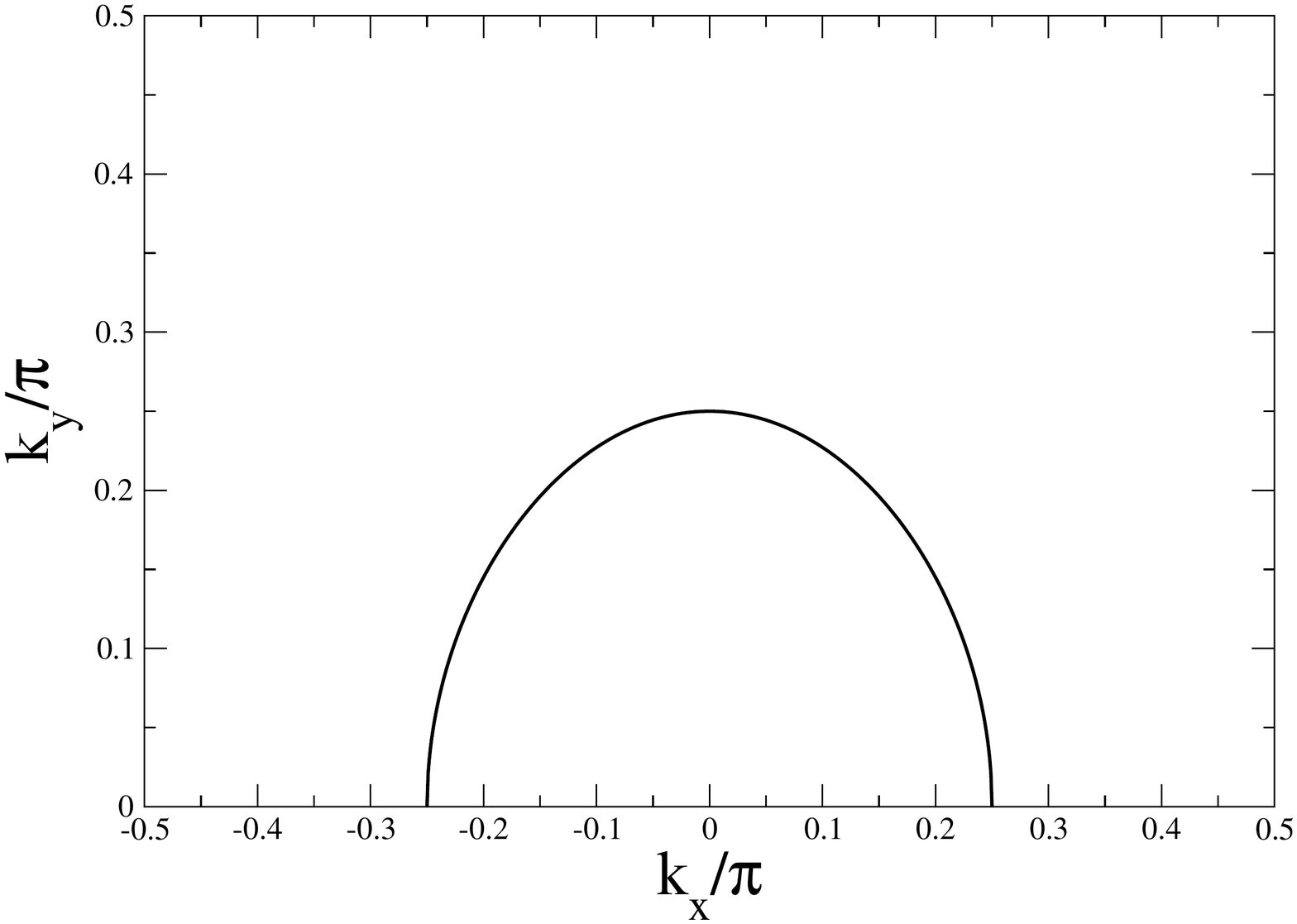}}
\caption{ The region of the surface BZ where Fermi arcs with $E>0$ on
  the bottom surface of a semi-infinite slab are allowed in our
  model. The parameters we have chosen for this plot are
  $k_0=\frac{\pi}{4}$, and $t'=0.8$. There is an identical region for
  $k_y<0$ which corresponds to arcs with $E<0$, again for the bottom
  surface of a semi-infinite slab, which is not shown.  }
\label{FAregion}
\end{figure}

The region is illustrated in Fig. (\ref{FAregion}) for $k_0=\frac{\pi}{4}$. In our simple model the spinors for the bottom surface are always
\begin{equation}
  \Phi_i^{(b)}=\left(\begin{array}{c}
    1\\
    0
  \end{array}\right),
\end{equation}
The eigenstates are spin-polarized along the $\sigma_z$
direction. Recalling the rotation of the Pauli matrices implemented in
the beginning of this analysis, this corresponds to polarization along
the $\sigma_y$ direction in the original model Eq. (1).

Now consider the top surface of a semi-infinite slab. In the lattice
Hamiltonian the values of $n$ now lie in the range
$n=0,-1,-2,-3,\dots,-\infty$. Instead of the ansatz we made for the bottom surface,
Eq. (\ref{decay-ansatz-b}), we now make the ansatz
\begin{equation}
  |E,k_x,k_y\rangle=\sum\limits_{n=0}^{\infty}
  \sum\limits_{s=\uparrow,\downarrow}u^{-n}\Phi_{0s}|n,s\rangle.
\label{decay-ansatz-t}
\end{equation}
Since $n$ now goes to large negative values  we see that we again want values of the two
$u_i$ which satisfy $|u_i|<1$.

The condition for the bulk to satisfy the eigenvalue equation remains
the same as Eq. (\ref{bulk-condition}), and the four roots of the
characteristic equation remain the same. Since, in our model, we have
shown that each root occurs with its reciprocal, we simply choose, in
the appropriate region (Eq. (\ref{FA-region})) the two values of $u$ to be
\begin{equation}
  u^{(t)}_i=u^{(b)}_i.
\end{equation}
For the bottom surface the energy is
\begin{equation}
  E^{(t)}(k_x,k_y)=-2f_y=-2\sin{k_y},
  \label{Etop}
\end{equation}
and the spinor is
\begin{equation}
  \Phi^{(t)}=\left(\begin{array}{c}
    0\\
    1
  \end{array}\right).
\end{equation}
At $E=0$ the Fermi arcs join the projections of the Weyl points onto
the surface BZ. The Fermi arc state on the bottom surface has a
(group) velocity in the positive $y$-direction while that on the top
surface has a group velocity in the negative $y$-direction.

We finally comment on the comparison with the continuum approximation
described in Sec. I. Within the lattice approach, the eigenstates are
parametrized by the multiplicative factor $u$ which denotes a decay of
the amplitude within a single lattice constant along the $z$
direction. The continuum solutions Eq. (\ref{eq:Phi_lambda}) for
$z=a=1$ implies that, in the appropriate limit, one should recover
$u=e^{-\lambda}$.  In particular, the continuum approximation is valid
for $\lambda_\pm\ll 1$; from the expression
Eq. (\ref{eq:lambda_final}), this is obeyed for $t'\ll 1$. For
simplicity, we further consider a range of $k_x$ where $\tilde{f}_x\ll
t'\ll 1$ (so that $\lambda_\pm$ are real). We then recall
Eq. (\ref{eq:u_pm}) for $u^{(b)}_{\pm}$ (with $t'=\sin\phi$ and
$\tilde{f}_x=F_x-1$), and expand $-\ln\{u^{(b)}_{\pm}\}$ to leading
order in $t',\tilde{f}_x$, to find
\begin{equation}
-\ln\{u^{(b)}_{\pm}\}=-\ln\left\{{F_x}\pm\sqrt{{F_x}^2-1+(t')^2}\right\}+\ln\{1+t'\}\approx t'-\tilde{f}_x\mp \sqrt{2\tilde{f}_x+(t')^2}\approx \lambda_\mp \; .
\end{equation}
Noting that, additionally, the solutions for the energy $E$ and eigenvectors are found to be identical, the continuum limit is indeed recovered in the appropriate limit.

\subsection{Two slabs rotated by $\frac{\pi}{2}$}
\label{lattice-twoslabs}

The key premise of the main text is that when the top surface of one
slab of WSM is tunnel-coupled to the bottom surface of another WSM,
the Fermi arcs on the two surfaces will hybridize and reconstruct.
Generically, since at arbitrary angles of rotation of one surface with
respect to the other, there is no true periodicity, and thus no
surface BZ, this problem cannot be handled by strictly lattice
methods. However, one simple case that can be handled by lattice
methods is when two slabs of the same WSM (assumed to have a square
lattice in the $xy$-plane) are rotated with respect to each other by
$\frac{\pi}{2}$ and tunnel-coupled. In the following we will assume
that the tunnel-coupling between the two surfaces in the $z$-direction
is identical in form to the coupling in the bulk, but may differ in
magnitude. We will use $c_n,\ n=0,1,2,\dots,\infty$ for the operators
of the top slab, and $d_n,\ n=0,-1,-2,\dots,-\infty$ for the operators
of the bottom slab, suppressing the spin and $k_x,k_y$ labels. Bearing
in mind that we are rotating the top slab by $\frac{\pi}{2}$ the total
Hamiltonian is
\begin{eqnarray}
  H=&\sum\limits_{n=0}^{-\infty} \bigg(d^{\dagger}_n\big(2\sin{k_y}\sigma_z+2\sigma_x(2+\cos{k_0}-\cos{k_x}-\cos{k_y})\big)d_n-d^{\dagger}_{n}(\sigma_x-it'\sigma_y)d_{n-1}-d^{\dagger}_{n-1}(\sigma_x+it'\sigma_y)d_{n}\bigg)\nonumber\\
  &+\sum\limits_{n=0}^{\infty} \bigg(c^{\dagger}_n\big(-2\sin{k_x}\sigma_z+2\sigma_x(2+\cos{k_0}-\cos{k_x}-\cos{k_y})\big)c_n-c^{\dagger}_{n+1}(\sigma_x-it'\sigma_y)c_n-c^{\dagger}_{n}(\sigma_x+it'\sigma_y)c_{n+1}\bigg)\nonumber\\
  &-\kappa d^{\dagger}_0(\sigma_x+it'\sigma_y)c_0-\kappa c^{\dagger}_0(\sigma_x-it'\sigma_y)d_0.
\label{Hcoupled}
\end{eqnarray}
Here $\kappa$ parameterizes the strength of the tunnel-coupling
between the two surfaces. Note the appearance of $-2\sin{k_x}$ as the
coefficient of $\sigma_z$ in the site-diagonal term in the Hamiltonian
for the top slab. This is because the slab has been rotated by
$\frac{\pi}{2}$, leading to $k_x\to k_y,\ k_y\to -k_x$. To reduce notational complexity in what follows, we introduce the following definitions.
\begin{eqnarray}
  &f^{(t)}_y(k_x,k_y)=-\sin{k_x},\nonumber\\
  &f^{(b)}_y(k_x,k_y)=\sin{k_y},\\
  &F^{(t)}_x=F^{(b)}_x\equiv F_x=2+\cos{k_0}-\cos{k_x}-\cos{k_y}\nonumber,\\
  &g^{(t)}(u)=-\frac{u(E-f^{(t)}_y)}{u^2(1+t')-2uF_x+1-t'}\nonumber,\\
  &g^{(b)}(u)=-\frac{u(E-f^{(b)}_y)}{u^2(1-t')-2uF_x+1+t'}.\nonumber
\end{eqnarray}
For a given $E,k_x,k_y$ let us call the two appropriate roots in the
top slab $u^{(t)}_i$, and the two in the bottom slab
$u^{(b)}_i$. Generically these will not be the same because
$f^{(t)}_y$ and $f^{(b)}_y$ differ. Further defining
\begin{equation}
  g^{(t)}_i\equiv g^{(t)}(u^{(t)}_i),\ \ \ \ \ \ \ \ g^{(b)}_i\equiv g^{(b)}(u^{(b)}_i),
\end{equation}
we write the wavefunctions in the top and bottom slabs as
\begin{eqnarray}
  \Phi^{(t)}(n)=\alpha_1 \big(u^{(t)}_1\big)^n\left(\begin{array}{c}
    1\\
     g^{(t)}_1\end{array}\right) +\alpha_2 \big(u^{(t)}_2\big)^n\left(\begin{array}{c}
    1\\
    g^{(t)}_2\end{array}\right),\\
  \Phi^{(b)}(n)=\beta_1 \big(u^{(b)}_1\big)^{-n}\left(\begin{array}{c}
    g^{(b)}_1\\
  1\end{array}\right) +\beta_2 \big(u^{(b)}_2\big)^{-n}\left(\begin{array}{c}
    g^{(t)}_2\\
  1\end{array}\right).
\end{eqnarray}
The boundary conditions at the interface can be compactly written as
\begin{equation}
  \Phi^{(t)}(n=-1)=\kappa\Phi^{(b)}(0),\ \ \ \ \ \ \ \Phi^{(b)}(n=+1)=\kappa\Phi^{(t)}(0)
  \label{BCpiover2}
\end{equation}
In explicit form, these four equations can be written as
\begin{equation}
  \left(\begin{array}{cccc}
    \big(u^{(t)}_1\big)^{-1}&\big(u^{(t)}_2\big)^{-1}&-\kappa g^{(b)}_1&-\kappa g^{(b)}_2\\
    g^{(t)}_1\big(u^{(t)}_1\big)^{-1}&g^{(t)}_2\big(u^{(t)}_2\big)^{-1}&-\kappa&-\kappa\\
    \kappa&\kappa&-g^{(b)}_1\big(u^{(b)}_1\big)^{-1}&-g^{(b)}_2\big(u^{(b)}_2\big)^{-1}\\
    \kappa g^{(t)}_1&\kappa g^{(t)}_2&-\big(u^{(b)}_1\big)^{-1}&-\big(u^{(b)}_2\big)^{-1}
  \end{array}\right)\left(\begin{array}{c}
    \alpha_1\\
    \alpha_2\\
    \beta_1\\
    \beta_2\end{array}\right)
    \equiv M
    \left(\begin{array}{c}
    \alpha_1\\
    \alpha_2\\
    \beta_1\\
    \beta_2\end{array}\right)
    =0
\end{equation}

Now suppose we want to find the reconstructed Fermi arc at the
interface for a given value of $E$. For a solution to exist the
determinant of the matrix $M$ must vanish. The only free
variables left in $M$ once $E$ is fixed are $k_x,k_y$, and $det(M)=0$
puts one condition on them. This is the implicit equation for the
reconstructed Fermi arc. In contrast to the results for the
semi-infinite slabs, the reconstructed Fermi arcs for $\kappa\neq 0$
do depend on the value of $t'$.

In Fig. (\ref{fig:Reconstruct-90}) we show the reconstructed Fermi arc
for $E=0$ for different values of the inter-slab coupling
$\kappa$. Only the first quadrant of the surface BZ is shown, with a
similar reconstruction occuring in the third quadrant.

\begin{figure}
\centerline{\includegraphics[angle=0,width=0.4\columnwidth]{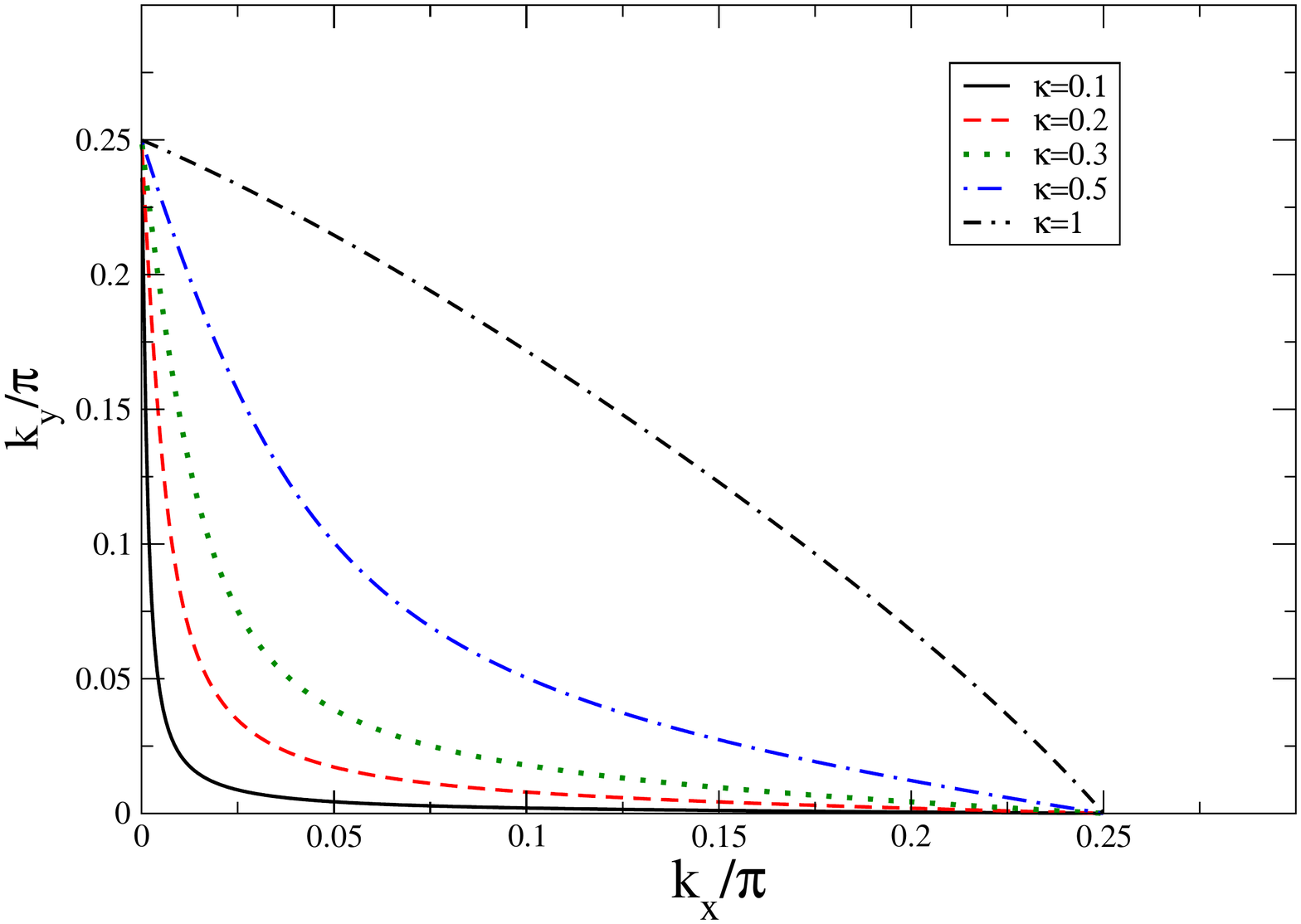}}
\caption{ Reconstructed Fermi arcs for two semi-infinite slabs of
  identical Weyl semimetals rotated by $\frac{\pi}{2}$ with respect to
  each other and the free surfaces tunnel-coupled with a strength
  $\kappa$. The parameters we have chosen for this figure are
  $k_0=\frac{\pi}{4}$ and $t'=\sin{\frac{\pi}{3}}$. Only the first
  quadrant is shown. There is a an identical reconstructed Fermi arc
  in the third quadrant as well. At zero tunnel-coupling the Fermi
  arcs are un-reconstructed. The Fermi arc belonging to the top slab
  will be the vertical line $k_x=0,\ -k_0<k_y<k_0$, while the Fermi
  arc belonging to the bottom slab will be the horizontal line
  $-k_0<k_x<k_0,\ k_y=0$. The projections of the Weyl points with
  positive monopole number are at $(0,k_0)$ (top slab) and $(k_0,0)$
  (bottom slab). As $\kappa$ increases, the surface states hybridize,
  and the Fermi arcs reconstruct to reconnect the projections of the +
  monopoles of the two slabs together and the - monopoles together. }
\label{fig:Reconstruct-90}
\end{figure}

\subsection{Curved Fermi arcs and detachment from Weyl points}
\label{curved-detach}

In the main manuscript we noted that at the arcless angles, when the
projections of the Weyl points of positive chirality coincide in the
surface Brillouin zone (and likewise for the negative chirality Weyl points), the
Fermi arcs may form closed loops which are detached from the Weyl points.
To demonstrate this requires a nonperturbative calculation in the
coupling between the two surfaces.
Analogous behavior has been demonstrated for $t'=1$ \cite{Diwedi_2018},
but the special properties of that parameter choice (discussed above)
leave open the question of whether such behavior is generic.
To address this, we present in this section an analogous calculation demonstrating
that Fermi loops can indeed detach from the projections of the
Weyl points for sufficiently strong tunnel-coupling, without fine-tuning of parameters.

We modify our model slightly to produce curved Fermi arcs, by changing
the functional form of $f_y$ to
\begin{equation}
  f_y(k_x,k_y;\lambda)=\sin{k_y}+\lambda(\cos{k_x}-\cos{k_0}).
  \label{fy-lambda}
\end{equation}
It is easy to see that the bulk Weyl points remain unchanged,
regardless of the value of $\lambda$.  All the manipulations of
Section \ref{lattice-oneslab} go through as before. From
Eq. (\ref{Ebottom}), for the bottom surface of a semi-infinite slab,
the Fermi arc at $E=0$ is given by
\begin{equation}
  E=2f_y(k_x,k_y;\lambda)=0.
\end{equation}
Let us choose the top semi-infinite slab as above with
$\lambda_{t}>0$, while the bottom slab has the same form with
$\lambda_{b}=-\lambda_{t}$. The two lattices are in registry with each
other, and the surfaces are brought into proximity, with the coupling
being of the same form as in Eq. (\ref{Hcoupled}). The formalism of
Section \ref{lattice-twoslabs} is general enough to accomodate this
situation with the appropriate replacements for
$f_y^{(t)},\ f_y^{(b)}$, so the implicit equation for the Fermi arcs
is once again given by the vanishing of the determinant of the
appropriate $M$ matrix.

In Fig. (\ref{liftoff}) we show the evolution of the Fermi arcs as the
coupling strength $\kappa$ between the two surfaces is
increased. There is a critical value of $\kappa^*\approx0.65$ beyond
which the Fermi arcs detach from the projections of the Weyl points
and form a closed loop detached from the Weyl point projections.

\begin{figure}
\centerline{\includegraphics[angle=0,width=0.4\columnwidth]{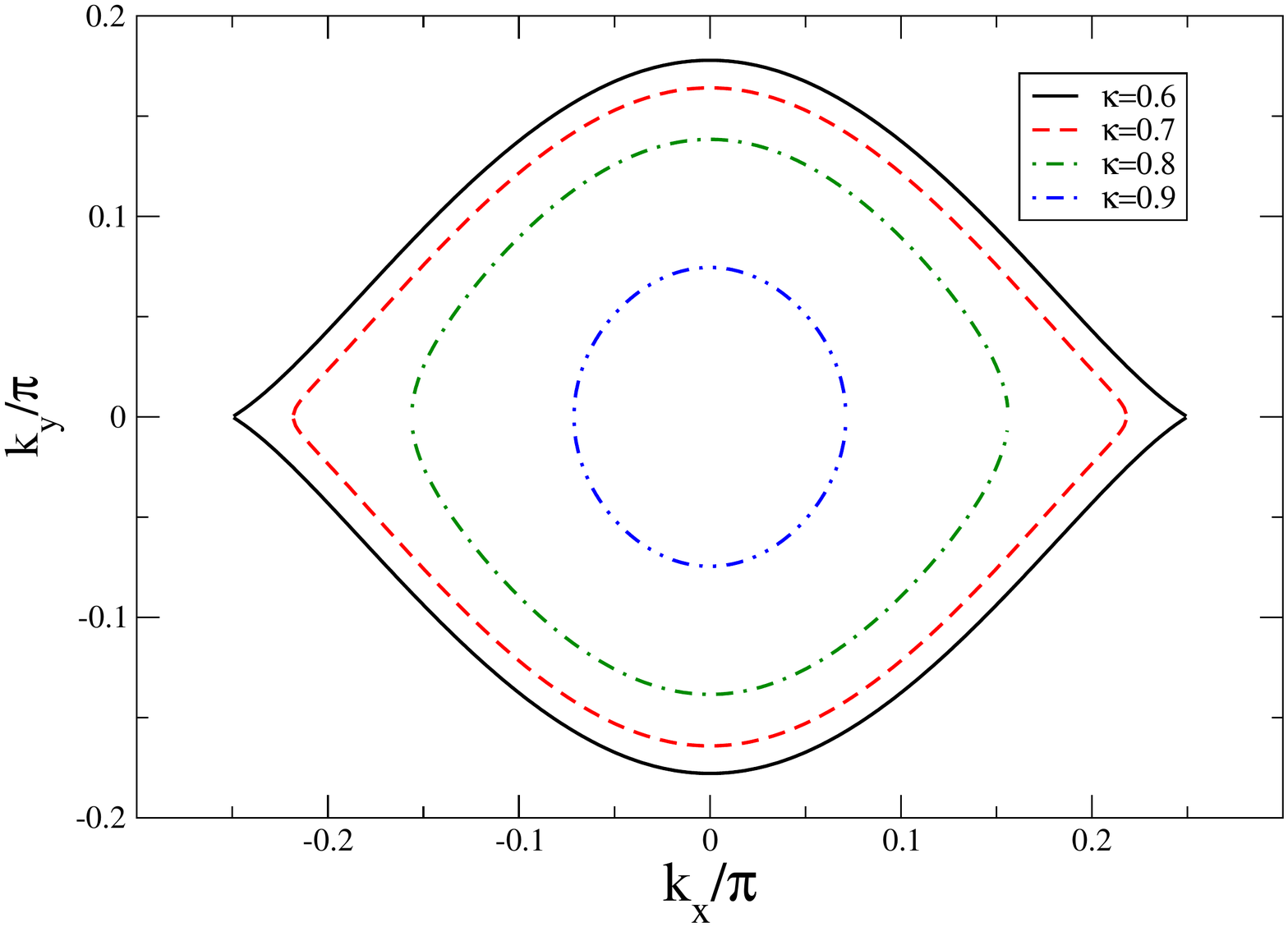}}
\caption{ Evolution of the Fermi arcs of the interface of two slabs
  with a tunnel-coupling $\kappa$ at the surface. Both slabs have bulk
  Weyl points at $(\pm k_0, 0,0)$. The values of the parameters are
  $k_0=\frac{\pi}{4}, t'=\sin\frac{\pi}{3}$. The top slab has
  $\lambda=2$ (see Eq. (\ref{fy-lambda})), while the bottom slab as
  $\lambda=-2$.  For small values of the tunnel-coupling $\kappa$, the
  Fermi arcs end at the projections of the Weyl points on the surface
  BZ. However, beyond a critical $\kappa^*\approx 0.65$ the Fermi arcs
  detach from the projections of the Weyl points and form an independent closed
  loop in the surface Brillouin zone.  This loop shrinks as $\kappa$ increases,
  and vanishes for $\kappa>1$.  }
\label{liftoff}
\end{figure}

\end{widetext}
\newpage

\bibliography{weyl_twist_ref2}

\begin{thebibliography}{31}
\bibitem[{foo()}]{foot1}
\bibinfo{note}{Note that these equations do not contain Berry's curvature
  terms, as the Fermi arc states are spin-polarized, and the phases of
  $w_{d,u}$ are not expected to have significant ${\bf k}$ dependence in the
  regions where Fermi arcs on opposite sides of the interface are degenerate.}
\end{thebibliography}

\bibliographystyle{apsrev}

\end{document}